\documentclass[12pt,a4paper]{article}

\usepackage{amsmath}
\usepackage{graphicx}
\usepackage{times}
\usepackage{cite}
\usepackage{ulem}
\begin{document}
\begin{titlepage}

\title{Energy dependence of deep-elastic scattering}
\author{ S.M. Troshin\footnote{Sergey.Troshin@ihep.ru}, N.E. Tyurin\\[1ex]
\small  \it NRC ``Kurchatov Institute''--IHEP\\
\small  \it Protvino, 142281, Russian Federation}
\normalsize
\date{}
\maketitle

\begin{abstract}
	
We discuss decreasing energy behavior of the differential cross-section 
of elastic hadron interactions at  fixed large values of  transferred 
momenta (deep elastic scattering). This observable effect is relevant 
to the reflective scattering mode.	
  
\end{abstract}
\end{titlepage}
\setcounter{page}{2}
\section*{Introduction}

The notion of  deep--elastic scattering was introduced in \cite{isl}. 
 Its relation \cite{twe} to scattering dynamics at sufficiently high energies can be associated 
 with antishadowing and its interpretation as a reflective scattering.
The  term anishadowing has been introduced in \cite{plb93} under consideration of the ``black disc  limit'' exceeding.    
This term means that reduction of the inelastic interactions contribution with the energy results in  the elastic amplitude increase. The respective scattering mode can take place in the limited range (dependent on the collision energy) of the impact parameter variation when 
$|f | > 1/2$ only, i.e. in the region where the amplitude is beyond  the so called black disc limit. This mode differs from the shadow case when both the elastic and inelastic  components of the unitarity equation  grow up with the energy increase, and $|f | <1/2$.

Explanation of antishadowing as a reflective scattering \cite{int2007} was proposed by analogy with optics. Appearance of the reflective scattering has been considered  along with  the positive reflective ability observation based on the LHC results \cite{tamas}. 
This  is not a unique interpretation of antishadowing. Indeed, an analogy with a resonant scattering was used  in \cite{anis,nekr}.  
We adhere to the reflective scattering  interpretation of antishadowing and use the respective appelation for it in what follows. 

Transition from  the shadow  to reflective scattering is developing with increase of the elastic scattering amplitude in the impact parameter representation\footnote{For the discussion of the impact parameter representation  see \cite{webb} and references therein.} with the energy growth when crossing the ``black disc limit''  \cite{plb24}.
  This evolution is connected with  formation of a peripheral impact parameter profile of the inelastic overlap function (black ring formation). The process of black ring formation starts at small impact parameter values. It is expected that this will manifest itself in  the elastic scattering differential cross--section behavior at large transferred momenta. The aim of this note is to trace an energy evolution of the relevant observable effect. 
  
  We use the $U$--matrix unitarization of the impact parameter dependent scattering amplitude and show in this note that its energy evolution leads to a decreasing energy dependence of the differential cross--section at fixed and large values of  the momentum transfer $-t$. 
  
  \section{Singularities  in the complex $\beta$-plane}
  The scattering amplitude $F(s,t)$ can be represented as a contour integral over the variable $\beta=b^2$ ($b$ is the impact parameter):
  \begin{equation}\label{con}
  F(s,t)=-\frac{is}{2\pi^2}\int_Cd\beta f(s,\beta) K_0(\sqrt{t\beta}),
  \end{equation}
where the contour $C$ encloses the positive semiaxis and finally, after its respective deformation,  the $F (s, t)$ is given \cite{tmf} as a sum of the contributions from the poles and cut of the amplitude 
	\begin{equation}\label{fsb}
		f(s,\beta)=u(s,\beta)/[1-iu(s,\beta)],
\end{equation} and 
  \begin{equation}
  	F(s,t)=F_p(s,t)+ F_c(s,t).
  \end{equation}

Singularities determining the amplitude behavior in the region of small and moderate momentum transfers are the poles in the complex $\beta$--plane. 
This kinematical region is associated with the function $F_p(s,t)$ \cite{plb24} .  Here we discuss the cut contribution $F_c(s,t)$ and its decisive importance for the large transferred momenta region.

The explicit form of the function $u$ is given  by the representation 
\begin{equation}\label{ml}
	u(s,\beta)=\frac{\pi^2}{s}\int_{t_0}^\infty\rho(s,x)K_0(\sqrt{x\beta})dx, 
\end{equation}
 where $\rho(s,x)$ is the spectral density which can be used for the model construction of $u$ and $t_0=4m^2$.

The  significant component of the dynamics is associated with  the input function $u$.
We  suppose that $u(s,\beta)$ monotonically increases with the energy in order to cover the whole range of the allowed values $|f|\leq 1$ and to provide  saturation of the unitarity limit  at $s\to\infty$. This saturation provides  reproduction of the principle of  maximal strength of strong interactions introduced by Chew and Frautchi \cite{chew}. 
It imples an existence of  the two modes and \it continuous \rm transition to the reflective scattering  starting  at $\beta=0$  at some energy value $s=s_r$. The reflective ability  becomes positive at  $s>s_r, \beta=0$.
Under quantitative analysis of the LHC data \cite{tamas},  evaluation   of $\sqrt{s_r}$  in $pp$--scattering was made and  the magnitude of $\sqrt{s_r}\simeq 13$ TeV has been obtained.

The scattering at sufficiently large impact parameters always remains to be in the shadow  mode. This region of impact parameters moves logarithmically with the energy to  periphery of the interaction region.
 Thus, the hadron scattering interaction region evolves   from a grey to black disk and then to a black ring with the reflective region in the center of the ring. Part of the amplitude $F_c(s,t)$ originating from the cut in the complex $\beta$ plane with a branching point at $\beta=0$ is mostly sensitive to the  energy evolution. We concentrate on the contribution $F_c(s,t)$ in the next section. 
\section{Differential cross--section at large transferred momenta }
Differential cross--section of the elastic scattering at large  transferred momenta is determined by the $F_c(s,t)$  since the function $F_p(s,t)$ decreases exponentially  in this region of $-t$. The contribution $F_c(s,t)$  can be represented in the form
\begin{equation}
F_c(s,t)=-\frac{s}{\pi^2}\int_{-\infty}^0 d\beta\, \mbox{disc}\, f(s,\beta)K_0(\sqrt{t\beta}), 
\end{equation}
where 
\begin{equation}
\mbox{disc} f(s,\beta)= \frac{\mbox{disc}\, u(s,\beta)}{[1-iu(s,\beta+i0)][1-iu(s,\beta-io)]}.
\end{equation}
Eq. (\ref{ml}) allows to get $\mbox{disc}\, u(s,\beta)$, i.e.
\begin{equation}\label{du}
\mbox{disc}\, u(s,\beta)= -\frac{\pi^3}{2s}\int_{t_0}^\infty dx \rho(s,x)J_0(\sqrt {x|\beta|}).
\end{equation}
On the base of Eq. (\ref{ml}), we assume for the function $u(s,\beta)$ the following simple expession ($\mu^2=t_0$)
\begin{equation}
	u(s,\beta)=ig(s)\exp [-\mu\sqrt{\beta}], 
\end{equation}
where the real function $g(s)\sim s^\lambda$, $\lambda >0$. Then, at sufficiently high energies and large values of $|t|$ ($|t|\gg \mu^2$), the differential cross--section of elastic scattering has the following dependence
\begin{equation}\label{dsig}
d\sigma/dt\sim g^{-2}(s)|t|^{-3}.	
\end{equation}

Eq. (\ref{dsig}) for $d\sigma/dt$ is valid for $g(s)\gg 1$, i.e. where the reflective scattering mode is assumed. Its presence results in the decreasing energy dependence of  the elastic scattering differential cross--section $d\sigma/dt \sim s^{-2\lambda}$ at fixed and large transferred momenta referred as deep--elastic scattering (DES) region \cite{isl}.

This prediction, Eq. (\ref{dsig}), is for the fixed transferred momenta (not scattering angles) and can, in principle, be checked experimentally. Unfortunately, there are no  data in the relevant kinematical region.  The available data \cite{dat, dat1} are in agreement with  decreasing energy dependence at fixed $-t$, but those have been obtained in rather modest energy range where  no indications have been found for the reflective scattering presence.

Note that the dependence   Eq. (\ref{dsig}) obeys to the Cerulus--Martin  lower bound \cite{cerm} formulated for the fixed scattering angles region on the base of Mandelstam analyticity. 
Eq. (\ref{dsig}) saturates the lower bound for the differential cross--section  when $s\to\infty$ and $-t$ is fixed and nonzero \cite{buon}. The latter is a consequence of a power growth of the input function $u(s,\beta)$ with energy which in turn is requred to provide saturation of the upper bound  of the scattering amplitude at $t=0$ in accordance with the principle of maximal strength in  strong interactions. This dynamically generated reciprocity emphasizes interconnection of the upper and lower bounds for  high energy scattering amplitude and represents an additional cause for further studies of the reflective scattering mode.

Differential cross--section   is factorized and depends  on the single variable $\tau$ ($\tau= g(s)^{2/3}|t|$) in the  reflective scattering region,  $d\sigma/dt\sim \tau^{-3}.$ 
It seems that such dependence in DES should correlate with the scaling  in elastic scattering at the LHC discussed recently in \cite{royon} since the  variables denoted by $\tau$ are functionally coincide  in both cases.

\section{Conclusion}
The  LHC experiments have indicated crossing of the so called black disc limit $|f|=1/2$ by the elastic scattering amplitude \cite{tamas}. This is a clear signal  of the reflective mode appearance which starts in central collisions, and the $U$--matrix unitarization continuously reproduces the transition from the shadow to reflective mode.  Appearance of this mode and its gradual evolvement  may be interpreted as a formation of a color conducting state of QCD matter  in the transient state under hadron collision \cite {jpg}. 

To study further  transition to the reflective scattering, it seems logical  to perform an energy scan of the differential cross--section of elastic scattering at fixed and large values of the momentum transfer $-t$  in the LHC energy range (DES domain). It was argued that  the reflective scattering mode  presence may be revealed by observation of a decreasing energy dependence of this cross-section.

An anticipated  interconnection of the upper and lower bounds for high energy 
scattering amplitude observed here  on the basis of a particular model 
parametrization may have a much more general meaning.

\end{document}